\documentclass[sigconf]{acmart}
\usepackage{tabularx}
\usepackage{pdfpages}
\usepackage{caption}
\usepackage{listings}
\usepackage{xcolor}

\definecolor{codegreen}{rgb}{0,0.6,0}
\definecolor{codegray}{rgb}{0.5,0.5,0.5}
\definecolor{codepurple}{rgb}{0.58,0,0.82}
\definecolor{backcolour}{rgb}{0.95,0.95,0.92}

\lstdefinestyle{mystyle}{
    backgroundcolor=\color{backcolour},   
    commentstyle=\color{codegreen},
    keywordstyle=\color{magenta},
    numberstyle=\tiny\color{codegray},
    stringstyle=\color{codepurple},
    basicstyle=\ttfamily\footnotesize,
    breakatwhitespace=false,         
    breaklines=true,                 
    captionpos=b,                    
    keepspaces=true,                 
    numbers=left,                    
    numbersep=5pt,                  
    showspaces=false,                
    showstringspaces=false,
    showtabs=false,                  
    tabsize=2
}
\AtBeginDocument{%
  \providecommand\BibTeX{{%
    \normalfont B\kern-0.5em{\scshape i\kern-0.25em b}\kern-0.8em\TeX}}}

\setcopyright{none}
\renewcommand\footnotetextcopyrightpermission[1]{} 

\begin{document}

\title{Toward Real-time Analysis of Experimental Science Workloads on Geographically Distributed Supercomputers}

\author{Michael Salim}
\email{msalim@anl.gov}
\affiliation{%
  \institution{Argonne National Laboratory}
  \streetaddress{9700 S Cass Ave}
  \city{Lemont}
  \state{Illinois}
  \country{USA}
  \postcode{60439}
}

\author{Thomas Uram}
\email{turam@anl.gov}
\affiliation{%
  \institution{Argonne National Laboratory}
  \streetaddress{9700 S Cass Ave}
  \city{Lemont}
  \state{Illinois}
  \country{USA}
  \postcode{60439}
}

\author{J. Taylor Childers}
\email{jchilders@anl.gov}
\affiliation{%
  \institution{Argonne National Laboratory}
  \streetaddress{9700 S Cass Ave}
  \city{Lemont}
  \state{Illinois}
  \country{USA}
  \postcode{60439}
}

\author{Venkat Vishwanath}
\email{venkat@anl.gov}
\affiliation{%
  \institution{Argonne National Laboratory}
  \streetaddress{9700 S Cass Ave}
  \city{Lemont}
  \state{Illinois}
  \country{USA}
  \postcode{60439}
}

\author{Michael E. Papka}
\email{papka@anl.gov}
\affiliation{%
  \institution{Argonne National Laboratory}
  \streetaddress{9700 S Cass Ave}
  \city{Lemont}
  \state{Illinois}
  \country{USA}
  \postcode{60439}
}
\affiliation{%
  \institution{Department of Computer Science, Northern Illinois University}
  \streetaddress{1425 W. Lincoln Hwy}
  \city{DeKalb}
  \state{Illinois}
  \country{USA}
  \postcode{60115}
}

\renewcommand{\shortauthors}{Salim, et al.}

\begin{abstract}
 Massive upgrades to science infrastructure are driving data velocities upwards while stimulating adoption of increasingly data-intensive analytics.  While next-generation exascale supercomputers promise strong support for I/O-intensive workflows, HPC remains largely untapped by live experiments, because data transfers and disparate batch-queueing policies are prohibitive when faced with scarce instrument time.  To bridge this divide, we introduce Balsam: a distributed orchestration platform enabling workflows at the edge to securely and efficiently trigger analytics tasks across a user-managed federation of HPC execution sites.  We describe the architecture of the Balsam service, which provides a workflow management API, and distributed sites that provision resources and schedule scalable, fault-tolerant execution.  We demonstrate Balsam in efficiently scaling real-time analytics from two DOE light sources simultaneously onto three supercomputers (Theta, Summit, and Cori), while maintaining low overheads for on-demand computing, and providing a Python library for seamless integration with existing ecosystems of data analysis tools.
\end{abstract}

\maketitle

\section{Introduction}

Department of Energy (DOE) supercomputers have been successfully leveraged by scientists for large-scale simulation in diverse fields, from understanding the evolution of the universe to discovering new materials. With the rise of data-intensive science, this landscape is changing to include high-throughput computing (HTC) workloads, typically comprising a large collection of jobs with a mix of interdependencies that govern the concurrency available to be exploited. A leading cause of this paradigm shift is experimental science: a range of experimental facilities are undergoing upgrades, or have recently completed upgrades, that will increase their data-taking rate, in some cases by as much as two orders of magnitude~\cite{osti_1398463, osti_1369223, osti_1341721}. This change necessitates access to computing resources beyond the experimental facility~\cite{superfacility,hepcloud}, to accommodate the real-time demands of experimental data analysis. Such workloads have been explored in a variety of past workflow systems~\cite{Atkinson2017}, job description approaches~\cite{Stubbs2020}, and methods of mapping jobs to compute resources~\cite{burkat2020serverless}.
Computational workloads from experimental facilities are uniquely driven by their coupling to ongoing experiments, leading to these features: \textit{significant data rates} that vary over time, from constant to bursty; the need for \textit{near real-time processing} to guide researchers operating the experiment to maximize the scientific utility of the instrument.  For instance, to probe nanoscale dynamics in materials such as amorphous ice, X-ray photon-correlation spectroscopy (XPCS) experiments collect frames of X-ray speckle pattern data on cycles of minutes or tens of minutes~\cite{Perakis8193}, evaluate the fidelity of the acquired time series by analyzing auto-correlation of pixelwise intensities, and perform repeated measurements across the experimental parameter space (e.g.\ sample position, temperature, pressure).  Today, XPCS detectors routinely collect megapixel frames at rates of 60 Hz ($\sim$ 120 MB/sec), but with recent advances in detector technology, frame rates of 11.8 kHz ($\sim$ 24 GB/sec) \cite{nsls-ii}  exacerbate the challenge of real-time analysis:  with limited computing resources, turnaround times of days or weeks preclude human-in-the-loop decision making or data-driven exploration of the parameter space.
Addressing this scientific bottleneck requires a multi-faceted approach to automated data reduction and analytics \cite{nsls-ii}, as well as  infrastructure to support the demands of data movement, storage, and large-scale computing and shrink analysis times from days or weeks to minutes.  

In the regime of \emph{near-real-time} analysis, where turnaround times on the order of 1-10 minutes are short enough to restore human-in-the-loop decision making for XPCS and similar experiments, a promising approach consists in distributing analysis workloads from the experiment across multiple remote computing facilities to avoid excessive batch queueing delays or (un)planned computing outages~\cite{osti_1341721,aps-upgrade, lcls-strategic-plan, als-strategic-plan}.  Our development of Balsam aims to support this approach with implementation in a few key areas: an application programming interface (API) for remote access to supercomputing facilities; close interaction with job schedulers to inject jobs opportunistically and bundle as ensembles to maximize throughput; interaction with data transfer mechanisms to move input (output) data to (from) remote locations, bundling transfers as needed; user-domain deployment, simplifying deployment while adhering to security requirements; and, distribution of jobs across multiple compute facilities simultaneously, with direction to a specific facility based on job throughput.

We demonstrate this functionality in the context of jobs flowing from two experimental facilities, the Advanced Photon Source (APS) at Argonne National Laboratory and the Advanced Light Source (ALS) at Lawrence Berkeley Laboratory, to three DOE compute facilities, the Argonne Leadership Computing Facility (ALCF), the Oak Ridge Leadership Computing Facility (OLCF), and NERSC. Balsam \textit{sites} were deployed at each facility to bridge the site scheduler to the Balsam service. The Balsam service exposes an API for introducing jobs and acts as a distribution point to the registered Balsam sites. Using the API, jobs were injected from the light source facilities simultaneously, executed at the computing sites, and results returned to the originating facility. Monitoring capabilities provide insight into the backlog at each Balsam site and, therefore, a basis for making scheduling decisions across compute facilities to optimize throughput.  While the current work emphasizes light source workloads, Balsam is easy to adopt and agnostic to applications, making this approach applicable to diverse scientific domains.

\vspace{-1em}
\section{Related Work \label{sec:related}}

Workflow management framework (WMF) development has a rich history and remains an active field of research and development. As an open-source Python WMF, Balsam~\cite{balsam-repo} is similar to other systems with Python-driven interfaces for managing high-throughput workflows on supercomputers.  On the other hand, some unique features of Balsam arise from requirements in orchestrating distributed workloads submitted from remote users and run on multiple HPC sites.  We now discuss these requirements and contrast Balsam with some contemporary HPC-oriented WMFs with Python APIs and distributed execution capabilities.


Distributed science workflows must invariably handle data and control flow across networks spanning administrative domains.  
Challenging barriers to WMF adoption in multi-user HPC environments lie in deploying distributed client/server infrastructures and establishing connectivity among remote systems. 
For instance, \textbf{Fireworks}~\cite{fireworks2015}, \textbf{Balsam}~\cite{xloops2019}, and the \textbf{RADICAL-Pilot / Ensemble Toolkit (EnTK)}~\cite{radical-pilot, radical2016} are three widely-used WMFs at DOE supercomputing facilities that expose Python APIs to define and submit directed acyclic graphs (DAGs) of stateful tasks to a database.  
Like Balsam, these WMFs possess various implementations of a common \emph{pilot job} design, whereby tasks are efficiently executed on HPC resources by a process that synchronizes state with the workflow database.  
Because the database is directly written by user or pilot job clients, users of Fireworks, RADICAL, or Balsam typically deploy and manage their own MongoDB servers \cite{mongodb} (plus RabbitMQ for EnTK, or PostgreSQL \cite{postgres} for Balsam).  The server deployments are repeated on a per-project basis and require provisioning resources on ``gateway'' nodes of an HPC system's local network.  Next, connecting to the database for remote job submission requires an approach such as SSH port forwarding \cite{xloops2019} or passwordless SSH/GSISSH \cite{radical-pilot}.  These methods are non-portable and depend on factors such as whether the compute facility mandates multi-factor authentication (MFA).

By contrast, the Balsam service-oriented architecture shifts administrative burdens away from individual researchers by routing all user, user agent, and pilot job client interactions through a hosted, multi-tenant web service.  Because Balsam execution sites communicate with the central service only as HTTP clients, deployments become a simple user-space \texttt{pip} package installation on any platform with a modern Python installation and outbound internet access.  As a consequence of the ubiquity of HTTP data transport, Balsam works ``out of the box'' on supercomputers spanning the ALCF, OLCF, and NERSC facilities against a cloud-hosted Balsam service.  


The Balsam REST API \cite{balsam-api} defines the unifying data model and service interactions, upon which all Balsam components and user workflows are authored.  
In the context of REST interfaces to HPC resources, Balsam may be compared with the \emph{Superfacility} concept \cite{superfacility}, which envisions a future of automated  experimental and observational science workflows linked to national computing resources through a set of Web APIs.  
Implementations such as the NERSC \textbf{Superfacility API} \cite{nersc-api} and the Swiss National Supercomputing Centre's \textbf{FirecREST API} \cite{firecrest} expose methods such as submitting and monitoring batch jobs, moving data between remote filesystems, and checking system availability.  
However, these facility services alone do not address workflow management or high-throughput execution; instead, they provide web-interfaced abstractions of the underlying facility, analogous to modern cloud storage and infrastructure services.  By contrast, Balsam provides an end-to-end, high-throughput workflow service, with APIs for efficiently submitting fine-grained tasks (as opposed to batch job scripts) along with their execution and data dependencies.  
Thus from a front-end view, Balsam may be viewed as an implementation of the Workflow as a Service (WFaaS) \cite{WFaaS} concept, with a backend architecture adapted to shared HPC environments.  

\textbf{funcX}~\cite{funcx2020} is an HPC-oriented instantiation of the function-as-a-service (FaaS) paradigm, where users invoke Python functions in remote containers via an API web service.  funcX endpoints, which are similar to Balsam \emph{sites} (\ref{sec:site}), run on the login nodes of target HPC resources where Globus Auth is used for authentication and user-endpoint association. 
The funcX model is tightly focused on Python functions and therefore advocates decomposing workflows into functional building blocks.  Each function executes in a containerized worker process running on one of the HPC compute nodes.  By contrast, Balsam provides a generalized model of applications, with executables that may or may not leverage containers, per-task remote data dependencies, programmable error- and timeout-handling, and flexible per-task resource requirements (e.g. tasks may occupy a single core or specify a multi-node, distributed memory job with some number of GPU accelerators per MPI rank).  In the context of end-to-end distributed science workflows, funcX is used as one tool in a suite of complementary technologies, such as \textbf{Globus Automate} \cite{automate-serverless} to set up event-based triggers of Globus data transfers and funcX invocations.  Balsam instead provides a unified API for describing distributed workflows, and the user site agents manage the full lifecycle of data and control flow. 


Implicit dataflow programming models, such as \textbf{Parsl}~\cite{parsl2019} and \textbf{Dask}~\cite{dask2016}, are another powerful option for authoring dynamic workflows directly in the Python language.  These libraries leverage the concept of Python futures to expose parallelism in the Python script and distribute computations across HPC resources.  In this context, the ephemeral workflow state resides in the executing Python program; tasks are not tied to an external persistent database.  This has obvious implications for provenance and fault tolerance of workflows, where per-workflow checkpointing is necessary in systems such as Parsl.  The Balsam service provides durable task storage for each user site, so that workflows can be centrally added, modified, and tracked over the span of a long-term project.  

\section{Implementation}
\begin{figure}
    \centering
    \vspace{-1em}
    \includegraphics[width=0.38\textwidth]{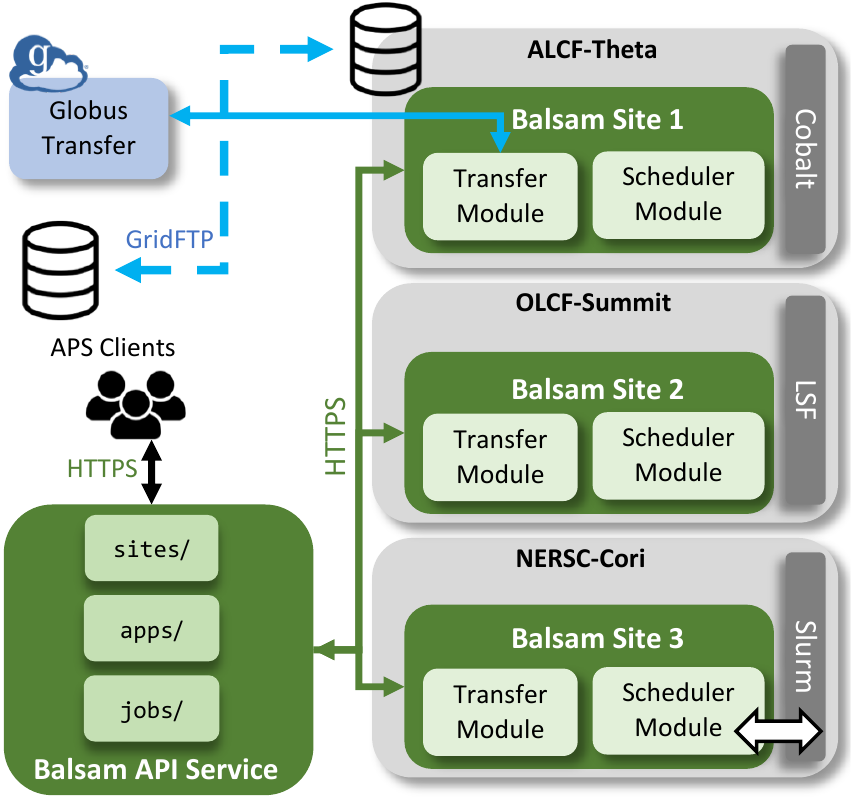}
    \caption{Simplified Balsam architecture.  Clients at the Advanced Photon Source (APS) submit scientific workloads  by creating \texttt{Job} resources at a specific execution site through the Balsam REST API \cite{balsam-api}.  Pilot jobs at Balsam sites on the ALCF-Theta, OLCF-Summit, and NERSC-Cori supercomputers fetch the appropriate \texttt{Jobs} for execution. All Balsam components independently communicate with the API service endpoints as HTTPS clients.  For instance, the Transfer Module fetches pending \texttt{TransferItems} from the API, bundles the corresponding files, and orchestrates out-of-band data transfers via Globus; the Scheduler Module synchronizes API \texttt{BatchJob} resources with the user's resource allocations on the local batch scheduler.}
    \label{fig:schematic-arch}
     \vspace{-1em}
\end{figure}

Balsam \cite{balsam-repo} (Figure \ref{fig:schematic-arch}) provides a centralized service allowing users to register execution \emph{sites} from any laptop, cluster, or supercomputer on which they wish to invoke computation.  The sites run user agents that orchestrate workflows locally and periodically synchronize state with the central service. Users trigger remote computations by submitting jobs to the   REST API \cite{balsam-api}, specifying the Balsam execution site and location(s) of data to be ingested. 

\subsection{The Balsam Service}

Balsam \cite{balsam-repo} provides a centrally-hosted multi-tenant web service  implemented with the FastAPI \cite{fastapi} Python framework and PostgreSQL \cite{postgres} relational database backend. The service empowers HPC users to manage distributed Balsam sites and orchestrate workflows within or across sites.
The centralized architecture, as opposed to \emph{per-user} databases, enables remote Balsam sites to communicate with the service as HTTPS clients.  This client-driven pattern is effective in many HPC environments, where outbound internet connectivity is common (or easily enabled by proxy), but other channels or inbound connections are often restricted, requiring non-portable solutions and involvement of system administrators.
We now discuss concepts in the REST API (refer to OpenAPI schema online \cite{balsam-api}), which defines the complete set of interactions upon which Balsam is built.

 The \textbf{Balsam User} is the root entity in the Balsam relational data model (e.g.\ all Sites are associated with their owner's user ID), and the service uses JSON Web Tokens (JWT) to identify the authenticated user in each HTTP request. Access Tokens are issued by Balsam upon authentication with an external provider:    we have implemented generic Authorization Code and Device Code OAuth2 flows \cite{device-oauth2} permitting users to securely authenticate with any OAuth2 provider.  The Device Code flow enables secure login from browserless environments such as supercomputer login nodes. The Balsam Python client library and \texttt{balsam login} command line interface manage interactive login flows and store/retrieve access tokens in the user's \texttt{{\raise.17ex\hbox{$\scriptstyle\sim$}/.balsam}} configuration directory.
In the near future, we anticipate integrating the Balsam OAuth2 system with a federated identity provider, as laid out in the DOE distributed data and computing ecosystem (DCDE) plan \cite{DCDE-Shankar2019}.  


The \textbf{Balsam Site} is a user-owned endpoint for remote execution of workflows. Sites are uniquely identified by a hostname and path to a \emph{site directory} mounted on the host.  The site directory is a self-contained project space for the user, which stores configuration data, application definitions, and a sandboxed directory where remote datasets are staged into \emph{per-task} working directories prior to execution.  


\begin{lstlisting}[language=Python, numbers=none, label={listing:corr-appdef}, caption=XPCS-Eigen corr ApplicationDefinition class which we evaluate with benchmark datasets in Section 4.]
from balsam.site import ApplicationDefinition
class EigenCorr(ApplicationDefinition):
    """
    Runs XPCS-Eigen on an (H5, IMM) file pair from a remote location.
    """
    corr_exe = "/software/xpcs-eigen2/build/corr"
    command_template = f"{corr_exe} inp.h5 -imm inp.imm"
    environment_variables = {
        "HDF5_USE_FILE_LOCKING": "FALSE",
    }
    parameters = {}
    cleanup_files = ["*.hdf", "*.imm", "*.h5"]
    transfers = {
        "h5_in": {
            "required": True,
            "direction": "in",
            "local_path": "inp.h5",
            "description": "Input HDF5 file",
            "recursive": False,
        },
        "imm_in": {
            "required": True,
            "direction": "in",
            "local_path": "inp.imm",
            "description": "Input IMM file",
            "recursive": False,
        },
        "h5_out": {
            "required": True,
            "direction": "out",
            "local_path": "inp.h5",  # output is input, modified in-place
            "description": "Output H5 file",
            "recursive": False,
        },
    }
\end{lstlisting}

As part of its security model, Balsam does not permit direct injection of arbitrary commands or applications through its API.   Instead, users write \emph{ApplicationDefinition} classes in Python modules stored inside the site directory (see example Listing \ref{listing:corr-appdef}). 
These modules serve as flexible templates for the permissible workflows that may run at each Balsam site.  \textbf{Balsam Apps} registered with the API merely index the ApplicationDefinitions in each site with a 1:1 
correspondence.
The ApplicationDefinition provides a declarative definition of the application configuration via special attributes such as the \texttt{command\_template}, which is interpreted as a shell command where adjustable parameters are enclosed in double-curly braces.  
The \texttt{transfers} attribute defines named \emph{slots} for files or directories to be staged in or out prior to (or after) execution.  ApplicationDefinition metadata are automatically serialized and synchronized with the corresponding REST API \texttt{App} resource.  Thus, a site's Apps become \emph{discoverable} via the web API, but maliciously submitted App data does not impact the execution of local ApplicationDefinitions. 


A \textbf{Balsam Job}  (used interchangeably with task) represents a single invocation of a Balsam App at a particular site. The job contains application-specific parameters (such as command line arguments), resource requirements (like the number of compute nodes or MPI ranks per node), and locations of data to be staged in or out before/after app execution.   Jobs can also specify parent job dependencies, so that a user may submit DAGs of Balsam jobs to the API. Each job references a \emph{specific} App and is therefore transitively bound to run at a single site upon creation (\texttt{Job} $\rightarrow$ \texttt{App} $\rightarrow$ \texttt{Site}).  Jobs carry persistent states  (enumerated in the REST API documentation \cite{balsam-api}) and metadata events generated from events throughout their lifecycle.
Files or directories that need to be staged \emph{in} or \emph{out} are associated with each Balsam Job and therefore passed to the API during job creation calls.  However, the service tracks each of these \emph{TransferItems} as a standalone unit of transfer between some Balsam Site and remote destination (defined with a protocol-specific URI such as Globus endpoint ID and path).  


The Balsam job represents a fine-grained unit of work (task), while many tasks are executed inside a pilot job that efficiently scales on the HPC compute nodes.  The pilot job and its associated resource allocation is termed a \textbf{Balsam BatchJob}.  Balsam \texttt{BatchJobs} can be automatically created by the Site agent (autoscaling) or manually requested through the REST API.
By default, a \texttt{BatchJob} at a Balsam site continuously fetches any locally-runnable Jobs and efficiently ``packs'' them onto idle resources.  
Balsam \emph{launcher} pilot jobs establish an execution \textbf{Session} with the service, which requires sending a periodic heartbeat signal to maintain a lease on acquired jobs.  The session backend guarantees that concurrent launchers executing jobs from the same site do not acquire and process overlapping jobs.  
The Session further ensures that critical faults causing ungraceful launcher termination do not cause jobs to be locked in perpetuity: the stale heartbeat is detected by the service and affected jobs are reset to allow subsequent restarts. 

A major feature of Balsam is the Python SDK, which is heavily influenced by and syntactically mirrors a subset of the Django object relational model (ORM) \cite{django-orm}. Rather than generating SQL and interfacing to a database backend, the Balsam SDK generates RESTful calls to the service APIs, and lazily executes network requests through the underlying API client library.  The SDK provides a highly consistent interface to query, (bulk-)create, (bulk-)update, and delete the Balsam resources described in the preceding sections.  For instance, {\small \texttt{Job.objects.filter(tags={"experiment": "XPCS"}, state="FAILED")}} 
produces an iterable query of all Balsam Jobs that failed and contain the \texttt{experiment:XPCS} tag.  Upon iteration, the lower-level REST Client generates the \texttt{GET /jobs} HTTP request with appropriate query parameters.  The returned Balsam Jobs are deserialized as \texttt{Job} objects which can be mutated and synchronized with the API by calling their \texttt{save()} method.

\subsection{Balsam Site Architecture \label{sec:site}}

Each Balsam site runs an authenticated user agent that performs REST API calls to the central service to fetch jobs, orchestrate the workflow locally, and update the centralized state. 

Indeed, the Balsam service plays a largely passive bookkeeping role in the architecture, where actions are ultimately client-driven.  This design  is a largely practical choice made to fit the infrastructure and security constraints of current DOE computing facilities.  Operationally speaking, a user must (1) log in to a site by the standard SSH authentication route, (2) authenticate with the Balsam service to obtain an access token stored at the HPC site, and (3) initialize the Balsam site agent in a directory on the HPC storage system.  The Balsam site is therefore an ordinary user process and interfaces with the HPC system as such; the architecture does not require elevated privileges or integration with specialized gateway services.  Typically, the Balsam site consists of a few long-running lightweight processes on an HPC login node; the only requirements are a Python installation (version $\geq$ 3.7) and outbound internet access (i.e.\ the user is able to perform HTTP requests such as \texttt{curl https://api.github.com}). The Balsam launchers (pilot jobs) also fetch jobs directly from the REST API; therefore, outbound \texttt{curl} functionality is needed from the head node where batch job scripts begin executing.  These requirements are satisfied by default on the Summit@OLCF and Cori@NERSC machines; on Theta@ALCF, the default site configuration indicates that an HTTP proxy should be used in launcher environments.  

Site configurations comprise a YAML file and \emph{job template} shell script that wraps the invocation of the pilot job launcher.  These configurations are easily modified, and a collection of default configurations is included directly as part of the Balsam software. The YAML file configures the local Balsam site agent as a collection of independent modules that run as background processes on a host with gateway access to the site's filesystems and HPC scheduler (e.g. an HPC login node). These modules are Python components written with the Balsam SDK  (described in the previous section) which uses locally stored access tokens to authenticate and exchange data with the REST API services.

The Balsam modules are architected to avoid hardcoding platform-dependent assumptions and enable facile portability across HPC systems.  Indeed, the majority of Balsam component implementations are platform-independent, and interactions with the underlying diverse HPC fabrics are encapsulated in classes implementing uniform \emph{platform interfaces}.  Below we touch on a few of the key Balsam site modules and the underlying platform interfaces:


The Balsam \textbf{Transfer Module} queries the API service for pending 
\texttt{TransferItems}, which are files or directories that need to be staged \emph{in} or \emph{out} for a particular Balsam job.  The module batches transfer items between common endpoints and issues transfer tasks to the underlying transfer interface.  For instance, the Globus transfer interface enables seamless interoperation of Balsam with Globus Transfer for dispatching out-of-band transfer tasks.  The transfer module registers transfer metadata (e.g. Globus task UUIDs) with the Balsam API and polls the status of transfer tasks, synchronizing state with the API.   For example, to enable the Transfer Module with Globus interoperability, the user adds a local endpoint ID and a list of trusted \emph{remote} endpoint IDs to the YAML configuration.  The user can also specify the maximum number of concurrent transfer tasks, as well as the maximum transfer batch size, which is a critical feature for bundling many small files into a single GridFTP transfer operation.  The Transfer Module is entirely protocol-agnostic and adding new transfer interfaces entails implementing two methods to \emph{submit} an asynchronous transfer task with some collection of files and \emph{poll} the status of the transfer.


The \textbf{Scheduler Module} uses an underlying scheduler platform interface to \emph{query} local batch scheduler queues (e.g.\ \texttt{qstat}) and \emph{submit} resource allocations (\texttt{qsub}) for pilot jobs onto the queue.  Interfaces to Slurm, Cobalt and LSF are provided to support DOE Office of Science supercomputing systems at NERSC, ALCF, and OLCF, respectively. The interface provides the necessary abstractions for interacting with the scheduler via subprocess calls and enables one to easily support other schedulers.  

The Balsam Scheduler Module is then platform-agnostic and serves to synchronize API \texttt{BatchJobs} with the local resource manager.  We emphasize that this module does not consider \emph{when} or \emph{how many} resources are needed; instead, it provides a conduit for \texttt{BatchJobs} created in the Balsam service API to become concrete pilot job submissions in a local queue.  
The Scheduler Module YAML configuration requires only specifying the appropriate scheduler interface (e.g.\ Slurm), an API synchronization interval, the local queue policies (partitions), and the user's project allocations.  


A separate concern is automating queue submission (i.e.\ \texttt{BatchJob} creation): in real-time computing scenarios, users may wish to allow the site to automatically scale resources in response to a workload with changing resource requirements over time.  

The \textbf{Elastic Queue Module} allows users to flexibly control auto-scaling behavior at a particular Balsam site by configuring a number of settings in the YAML file: the appropriate scheduler queue, project allocation, and pilot job mode are specified, along with minimum/maximum numbers of compute nodes and minimum/maximum walltime limits.  The user further specifies the maximum number of auto-queued jobs, the maximum queueing wait time (after which \texttt{BatchJobs} are deleted from the queue), and a flag indicating whether allocations should be constrained to \emph{idle} (backfill) node-hour windows.  

At every sync period (set in the YAML configuration), the Elastic Queue module queries the Balsam API for the aggregate resource footprint of all runnable \texttt{Jobs}  (\emph{how many nodes could I use right now}) as well as the aggregate size of all queued and running \texttt{BatchJobs} (\emph{how many nodes have I currently requested or am I running on}).  If the runnable task footprint exceeds the current \texttt{BatchJob} footprint, a new \texttt{BatchJob} is created to satisfy the YAML constraints as well as the current resource requirements.


The Elastic Queue Module's backfill mode enables remote users to automatically tap unused cycles on HPC systems, thereby increasing overall system utilization, which is a key metric for DOE leadership-class facilities, with minimal perturbations to the scheduling flow of larger-scale jobs.  
By and large, other mechanisms for ensuring quality-of-service to experimentalists are dictated by DOE facility policies and therefore tangential to the current work.  However, we expect that Balsam autoscaling can seamlessly utilize on-demand or preemptible queues such as the \emph{realtime} queue  offered on Cori at NERSC.


The \textbf{Balsam launcher} is a general pilot job mechanism that enables efficient and fault-tolerant execution of Balsam jobs across the resources allocated within an HPC batch job. Launchers rely on two platform interfaces: the \emph{ComputeNode} interface defines available cores, GPUs, and MAPN (multiple applications per node) capabilities for the platform's compute nodes.  
The interfaces also includes method for detecting available compute nodes within an allocation and logically tracking assignment of granular tasks to resources. 
The \emph{AppRun} interface abstracts the application launcher, and is used to execute applications in an  MPI implementation-agnostic fashion.  


\section{Evaluation \label{sec:evaluation}}

In this section, we evaluate the performance characteristics of Balsam in remote workload execution and ability to dynamically distribute scientific workloads among three DOE supercomputing platforms.  We study two benchmark workflows in which a \emph{full round-trip analysis} is performed: data is first transferred from a client experimental facility to the supercomputer's parallel file system, computation on the staged data occurs on dynamically provisioned nodes, and results are finally transferred back from the supercomputer to the client's local file system.  

Our experiments address the following questions: (1) How effectively does scientific throughput scale given the real-world overheads in wide-area data transfers and in Balsam itself?  (2) Is the total round-trip latency consistently low enough to prove useful in \emph{near real-time} experimental analysis scenarios?  (3) Does Balsam complete workloads reliably under conditions where job submission rates exceed available computing capacity and/or resources become unavailable mid-run?  (4) Can \emph{simultaneously provisioned} resources across computing facilities be utilized effectively to distribute an experimental workload in real-time?  (5) If so, can the Balsam APIs be leveraged to adaptively map jobs onto distributed resources to improve overall throughput?

\subsection{Methodology}


\subsubsection{Systems}
\begin{figure}
    \centering
    \includegraphics[width=0.37\textwidth]{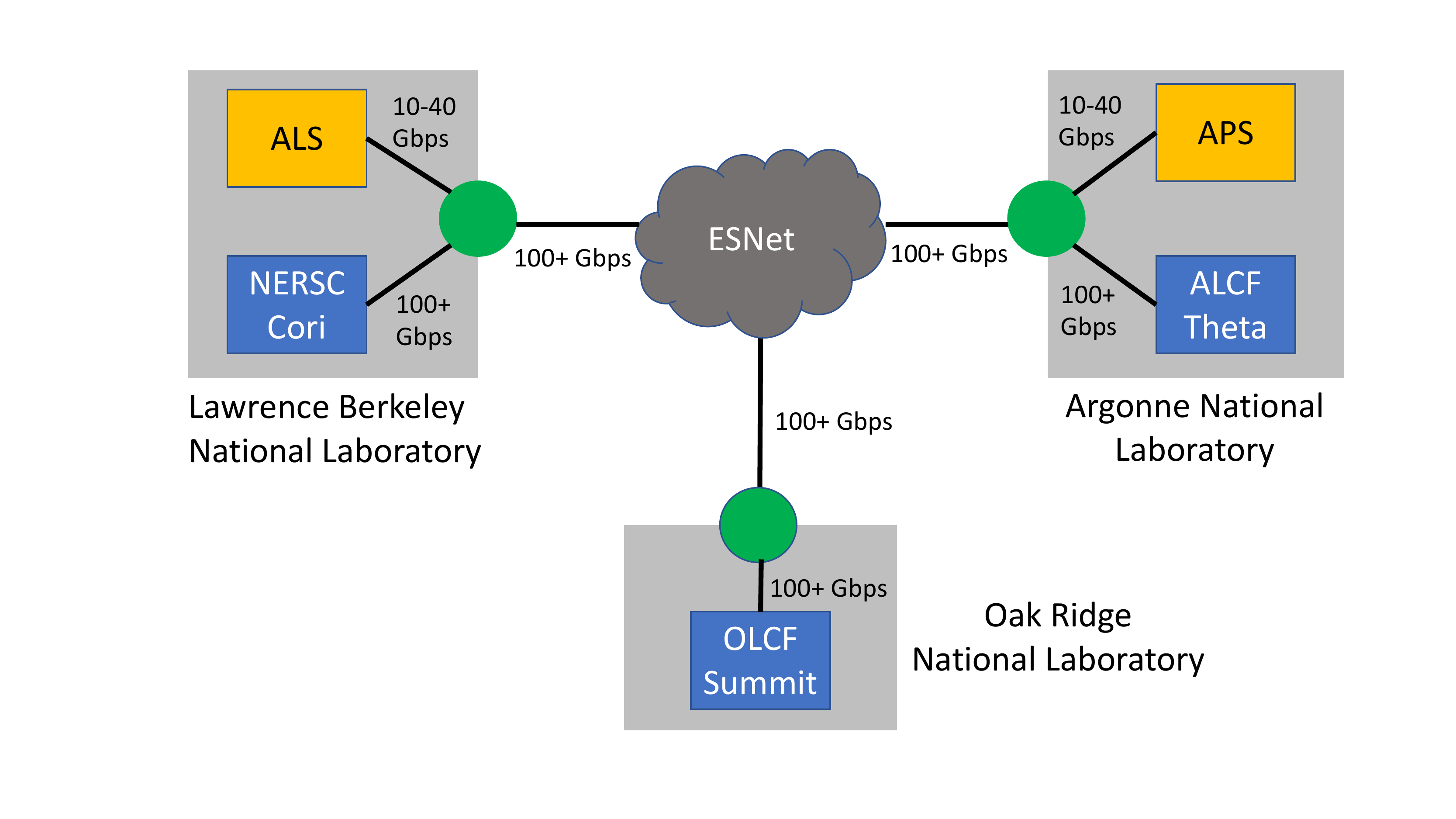}
    \vspace{-2em}
    \caption{Connectivity between the geographically distributed Balsam sites (Theta at ALCF, Cori at NERSC,  Summit at OLCF) and X-ray sources (APS at Argonne, ALS at LBNL). All networks are interconnected via ESNet.}
    \label{fig:network}
    \vspace{-2em}
\end{figure}

Figure \ref{fig:network} depicts the  geographically distributed supercomputing systems comprising of: 
a) \textbf{Theta~\cite{theta}:} A Cray XC40 system at the ALCF, with 4392 compute nodes, having 64 1.3 GHz Intel Xeon Phi (Knight's Landing) cores per node each with 192 GB memory.
b) \textbf{Summit~\cite{summit}:} An IBM AC922 system at the OLCF with 4,608 nodes, each node with two IBM Power 9 processors and six Nvidia V100 GPUs together with 512GB DDR4 and 96GB HBM2 memory.
c) \textbf{Cori (Haswell Partition)~\cite{cori}:} The Haswell partition of Cori, a Cray XC40 system at NERSC, consists of X nodes each with two 2.3 GHz 16-core Intel Haswell processor with 128 GB memory.  

The APS light source is located at Argonne and the ALS at LBNL. All three sites are interconnected via ESNet.



\subsubsection{Balsam}
The Balsam service, comprising both relational database and web server frontend, was deployed on a single AWS \texttt{t2.xlarge} instance. 
For the purpose of this experiment, user login endpoints were disabled and JWT authentication tokens were securely generated for each Balsam site.  Moreover, inbound connectivity was restricted to traffic from the ANL, ORNL, and NERSC campus networks.  

Three Balsam sites were deployed across the login nodes of Theta, Summit, and Cori.  Balsam managed data transfers via the Globus Online \cite{globus-online} interface, between each compute facility and data transfer nodes at the Argonne Advanced Photon Source (APS) and Berkeley Lab Advanced Light Source (ALS).  Data staging was performed with Globus endpoints running on dedicated data transfer nodes (DTNs) at each facility.  All application I/O utilized each system's production storage system (Lustre on Theta, IBM Spectrum Scale on Summit, and the project community filesystem (CFS) on Cori).  The ALCF internal HTTPS proxy was used to enable outbound connections from the Theta application launch nodes to the Balsam service.  Outbound HTTPS requests were directly supported from Cori and Summit compute nodes.

Since DOE supercomputers are shared and perpetually oversubscribed resources, batch queueing wait times are inevitable for real workflows.  
Although cross-facility reservation systems and \emph{realtime} queues offering a premium quality of service to experiments will play an important role in  real-time science on these systems, we seek to measure Balsam overheads in the absence of uncontrollable delays due to other users. In the following studies, we therefore reserved dedicated compute resources to more consistently measure the performance of the Balsam platform.

\subsubsection{Applications \label{sec:apps}}

We chose two applications to measure throughput between facilities in multiple scenarios. The \emph{matrix diagonalization} (MD) benchmark, which serves as a proxy for an arbitrary fine-grained numerical subroutine, entails a single Python NumPy call to the \texttt{eigh} function, which computes the eigenvalues and eigenvectors of a Hermitian matrix.  The input matrix is transferred over the network from a client experimental facility and read from storage; after computation, the returned eigenvalues are written back to disk and transferred back to the requesting client's local filesystem.  We performed the MD benchmark with double-precision matrix sizes of $5000^2$ (200 MB) and $12000^2$ (1.15 GB), which correspond to eigenvalue (diagonal matrix) sizes of 40 kB and 96 kB, respectively. These sizes are representative of typical datasets acquired by the XPCS experiments discussed next.

The \emph{X-ray photon correlation spectroscopy} (XPCS) technique presents a data-intensive workload in the study of nanoscale materials dynamics with coherent X-ray synchrotron beams, which is widely employed across synchrotron facilities such as the Advanced Photon Source (APS) at Argonne and the Advanced Light Source (ALS) at Lawrence Berkeley National Lab.  
 We chose to examine XPCS, because XPCS analysis exhibits a rapidly growing frame rates: with megapixel detectors and on the order of 1 MHz acqusition rates on the horizon, XPCS will push the boundaries of real-time analysis, data-reduction techniques, and networking infrastructure. \cite{aps-upgrade}
In this case, we execute  XPCS-Eigen \cite{xpcs-eigen} C++ package, used in the XPCS beamline, and the \texttt{corr} analysis  which computes pixel-wise time correlations of area detector images of speckle patterns produced by hard X-ray scattering in a sample.  The input comprises two files: a binary \texttt{IMM} file containing the sequence of acquired frames, and an \texttt{HDF} file containing experimental metadata and control parameters for the \texttt{corr} analysis routine.  \texttt{corr}  modifies the \texttt{HDF} file with results \emph{in-place} and the file is subsequently transferred back to the Balsam client facility.  
For the study, we use an XPCS dataset comprised of 823 MB of \texttt{IMM} frames and 55 MB of \texttt{HDF} metadata.  This 878 MB payload represents roughly 7 seconds of data acquisition at 120 MB/sec, which is a reasonably representative value for XPCS experiments with megapixel area detectors operating at 60 Hz and scanning samples for on the order of tens of seconds per position \cite{Perakis8193, nsls-ii}
All the XPCS and MD runs herein were executed on a single compute node (i.e. without distributed-memory parallelism) and leverages OpenMP threads scaling to physical cores per node (64 on Theta, 42 on Summit, 32 on Cori).

\subsubsection{Evaluation metrics}
The Balsam service stores Balsam Job events (e.g. \emph{job created}, \emph{data staged-in}, \emph{application launched}) with timestamps recorded at the job execution site. We leverage this Balsam \texttt{EventLog} API to query this information and aggregate a variety of performance metrics:

\begin{itemize}
    \item Throughput timelines for a particular job state (e.g. number of completed round-trip analyses as a function of time)
    \item Number of provisioned compute nodes and their instantaneous utilization (measured at the workflow level by number of running tasks)
    \item Latency incurred by each stage of the Balsam Job life-cycle
\end{itemize}

\subsubsection{Local cluster baselines at experimental beamlines \label{sec:local-cluster-def}}

For an experimental scientist, to understand the value of Balsam to offload analysis of locally acquired data, a key question to consider:  \emph{is the greater capacity of remotely distributed computing resources worth the cost of increased data transfer times and the time to solution?}  We therefore aim to measure baseline performance in a pipeline that is characteristic of  data analysis workflows executing on an HPC cluster near the data acquisition source  as performed in production at the beamlines today\cite{aps-dm}. In this case, the  imaging data is first copied from the local acquisition machine storage to the beamline cluster storage and analysis jobs are launched next to analyze the data. To that end, we simulated ``local cluster'' scenarios on the Theta and Cori supercomputers by submitting MD benchmark runs continuously onto an exclusive reservation of 32 compute nodes (or a fraction thereof) and relying on the batch scheduler, not Balsam, to queue and schedule runs onto available resources.  The reservations guaranteed resource availability during each run, so that measurements did not incur queuing delays due to other users on the system.  Instead of using Globus Online to trigger data movement, transfers were invoked in the batch job script as file copy operations between a data source directory and a temporary sandbox execution directory on the same parallel filesystem.  Thus, we consider the overall \emph{tradeoff} between increased data transfer time and reduced application startup time when adopting a remote WMF such as Balsam.


\subsection{Comparison to local cluster workflows}

\begin{figure*}
    \centering
    \includegraphics[width=0.8\textwidth]{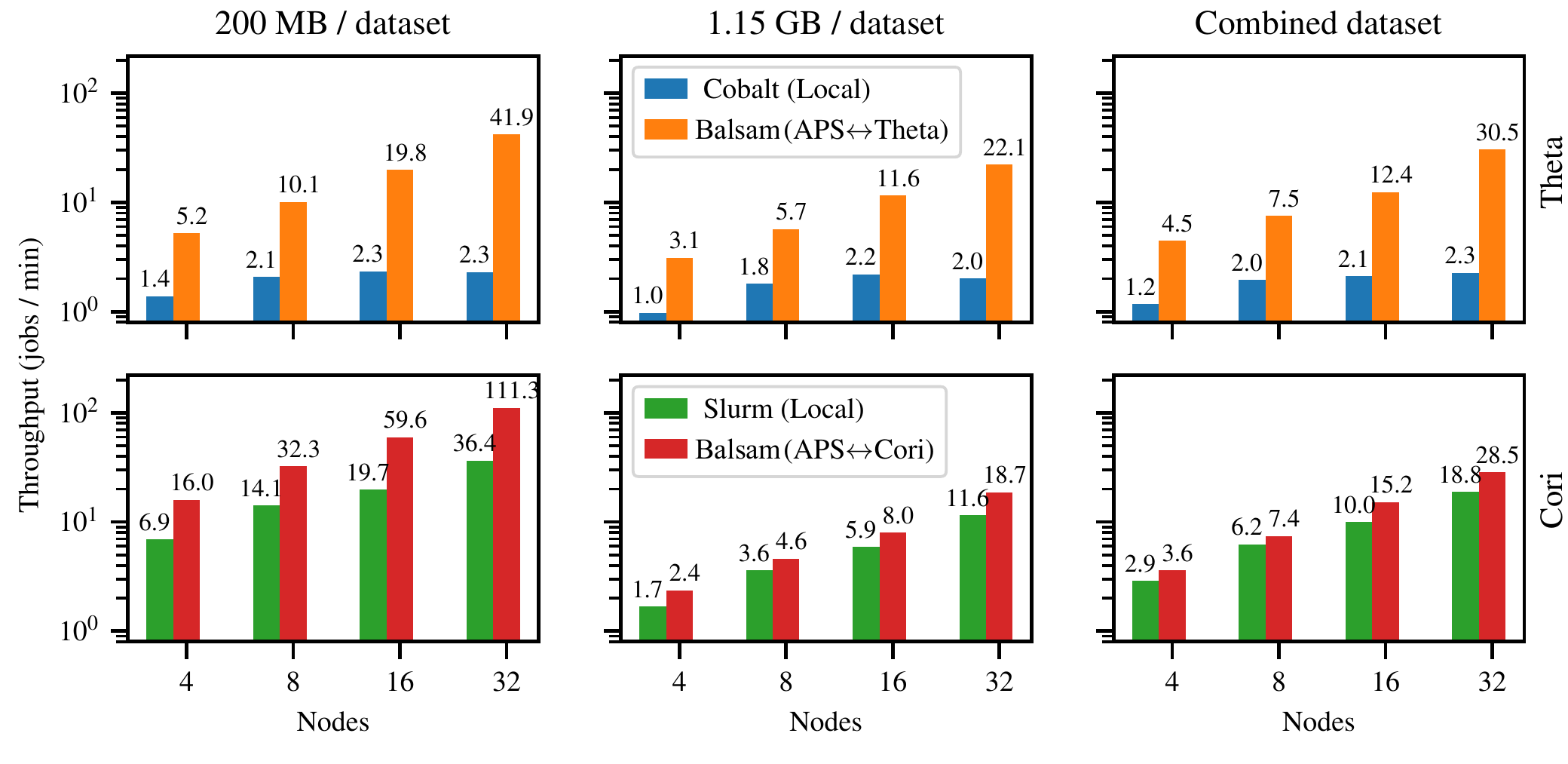}
    \vspace{-1.5em}
    \caption{Weak scaling of matrix diagonalization throughput on Theta (top row) and Cori (bottom).  Average rate of task completion is compared between local Batch Queue and Balsam APS $\leftrightarrow$ ALCF/NERSC pipelines  at various node counts.  The left and center panels show runs with $5000^2$ and $12000^2$ input sizes, respectively.  The right panels show runs where each task submission draws uniformly at random from the two input sizes.  To saturate Balsam data stage-in throughput, the job source throttled API submission to maintain steady-state backlog of up to 48 datasets in flight.  Up to 16 files were batched per Globus transfer task.}
     \vspace{-1.5em}
    \label{fig:throughput-matrix}
\end{figure*}

\begin{figure}
    \centering
    \includegraphics[width=0.4\textwidth]{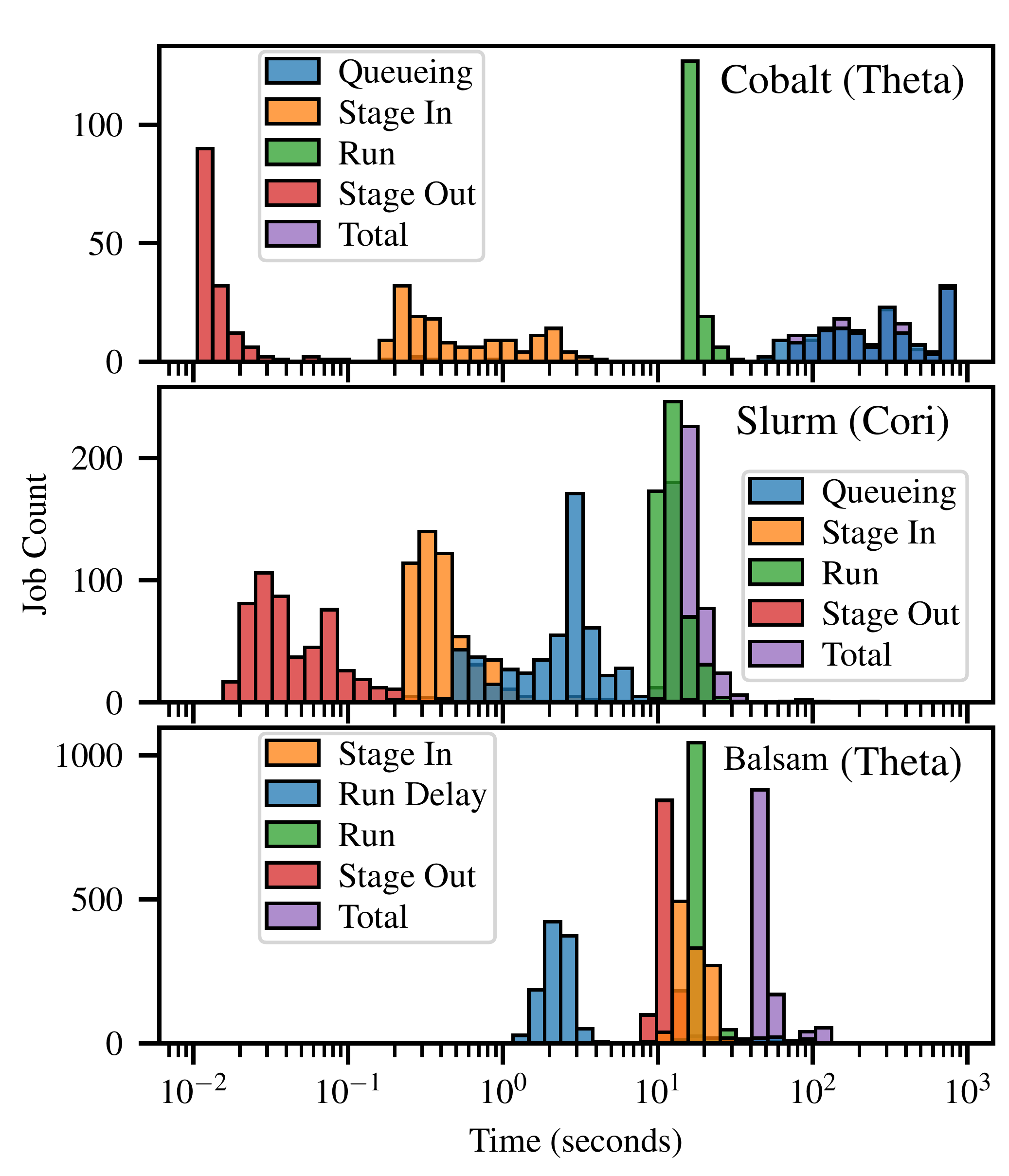}
     \vspace{-1.5em}
    \caption{Unnormalized histogram of latency contributions from stages in the Cobalt Batch Queuing (top), Slurm Batch Queuing (center), and APS $\leftrightarrow$ Theta Balsam (bottom) analysis pipelines in the 200 MB matrix diagonalization benchmark.  Jobs were submitted to Balsam API at a steady rate of 2 jobs/second onto a 32-node allocation.  
     The Queueing time represents delay between job submission to Slurm or Cobalt (using an exclusive node reservation) and application startup.  Because  Balsam jobs execute in a pilot launcher, they do not incur a queueing delay; instead, the \emph{Run Delay} shows the elapsed time between Globus data arrival and application startup within the pilot.
    \label{fig:balsam-vs-queues}
    }
     \vspace{-1.5em}
\end{figure}

Figure \ref{fig:throughput-matrix} depicts the scalability of Balsam task throughput when remotely executing the MD application on Theta or Cori with input datasets stored at the APS.  These measurements are compared with the throughput achieved by the ``local cluster'' pipeline, in which the same workload is repeatedly submitted onto an exclusive reservation of compute nodes via the local batch scheduler  as described in section \ref{sec:local-cluster-def}.  We consider both Theta,  which uses the Cobalt scheduler
and Cori, which uses the Slurm scheduler. 
Note that on both machines, the Balsam workflow entails staging matrix data in from the APS Globus endpoint and transferring the output eigenvalues in the reverse direction back to APS.  
 While the local cluster baseline does not use pilot jobs and is at a disadvantage in terms of inflated scheduler queueing times, the local runs simply read data from local storage and are also significantly \emph{advantaged} in terms of reduced transfer overhead.  This is borne out in the ``Stage In'' and ``Stage Out'' histograms of figure \ref{fig:balsam-vs-queues},  which show that local data movement times for the small (200 MB) MD benchmark are one to three orders of magnitude faster than the corresponding Globus transfers invoked by Balsam.

Despite added data transfer costs, Balsam efficiently overlaps asynchronous data movement with high-throughput task execution, yielding overall enhanced throughput and scalability relative to the local scheduler-driven workflows.  On Theta, throughput increases from 4 to 32 nodes with 85\% to 100\% efficiency, depending on the sizes of transferred input datasets.  On the other hand, the top panels of Figure \ref{fig:throughput-matrix} show that Cobalt-driven task throughput is non-scalable for the evaluated task durations on the order of 20 seconds (small input) or 1.5 minutes (large input).  The Cobalt pipeline is in effect throttled by the scheduler job startup rate, with a median per-job queuing time of 273 seconds despite an exclusive reservation of idle nodes.  In contrast to Cobalt on Theta, the bottom panels of Figure \ref{fig:throughput-matrix} show that Slurm on Cori is moderately scalable for high-throughput workloads.  For the smaller MD dataset, Slurm achieves 66\% efficiency in scaling from 4 to 32 nodes; for the larger dataset efficiency increases to 85\%.  The improvement over Cobalt is due to the significantly reduced median Slurm job queuing delay of 2.7 seconds (see blue ``Queueing'' duration histograms in figure \ref{fig:balsam-vs-queues}).  Here too, Balsam scales with higher efficiency than the local baseline (87\% for small dataset; 97\% for large dataset), despite the wide geographic separation between the APS and NERSC facilities. 

Provided adequate bandwidth, tapping remote resources clearly enables one to scale workloads beyond the capacity of local infrastructure, but this often comes at the price of unacceptably high turnaround time, with significant overheads in data movement and batch queuing. Table \ref{tab:balsam} quantifies the Balsam latency distributions that were measured from submitting 1156 small and 282 large MD tasks from APS to Theta at rates of 2.0 and 0.36 jobs/second, respectively.  We note a mean 34.1 second overhead in processing the small MD dataset and  $95^\textrm{th}$ percentile overheads of roughly 1 or 2 minutes in processing the small or large MD datasets, respectively.  On average, 84\% to 90\% of the overhead is due to data transfer time and not intrinsic to Balsam (which may, for instance, leverage direct memory-to-memory transfer methods).  Notwithstanding room for improvement in transfer rates, turnaround times on the order of 1 minute may prove acceptable in \emph{near-real-time} experimental scenarios requiring access to large-scale remote computing resources.   These latencies may be compared with other WMFs such as funcX \cite{funcx2020} and RADICAL-Pilot \cite{radical-pilot} as part of future work.

%

\begin{table}[]
\small
\caption{APS $\leftrightarrow$ Theta Balsam analysis pipeline stage durations (in sec) for the matrix diagonalization benchmark. The
latency incurred by each stage is reported as the mean $\pm$ standard deviation, with the 95\textsuperscript{th} percentile duration in parenthesis.  Jobs were submitted to Balsam API at a steady rate to a 32-node allocation (2.0 and 0.36 jobs/sec for small and large datasets respectively).} 
\label{tab:balsam}
\begin{tabular}{l||l|l}
\toprule
Stage &       200 MB (1156 runs) &  1.15 GB (282 runs)      \\
&       Duration (seconds) &  Duration (seconds)      \\
\midrule
Stage In             &    $17.1 \pm 3.8 \ (23.4)$    &    $47.2 \pm 17.9 \ (83.3)$ \\
Run Delay            &    $5.3 \pm 11.5 \ (37.1)$    &    $7.4 \pm 14.7 \ (44.6)$ \\
Run                  &    $18.6 \pm 9.6 \ (30.4)$    &    $89.1 \pm 3.8 \ (95.8)$ \\
Stage Out            &    $11.7 \pm 2.1 \ (14.9)$    &    $17.5 \pm 8.1 \ (34.1)$ \\
\midrule
Time to Solution     &    $52.7 \pm 17.6 \  (103.0)$ &  $161.1 \pm 23.8 \ (205.0)$ \\
Overhead             &    $34.1 \pm 12.3 \ (66.3)$   &   $72.1 \pm 22.5 \  (112.2)$ \\
\bottomrule
\end{tabular}
 \vspace{-1.5em}
\end{table}

\subsection{Optimizing the data staging}

\begin{figure}
    \centering
    \includegraphics[width=0.4\textwidth]{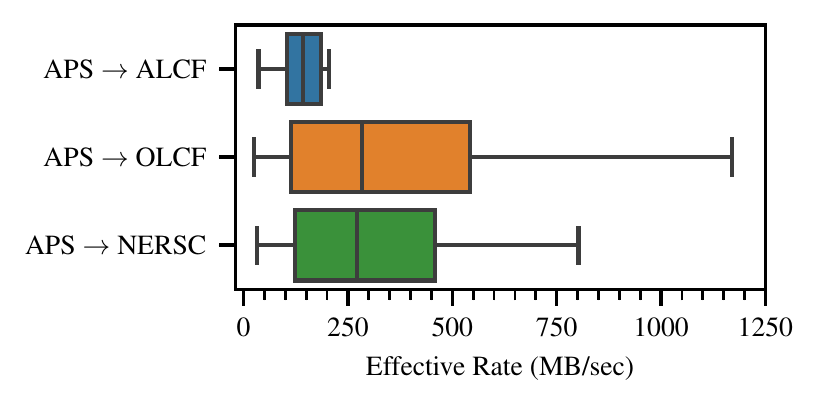}
     \vspace{-1.5em}
    \caption{Effective cross-facility Globus transfer rates. The transfer duration is measured from  initial API request to transfer completion; the rate factors in average transfer task queue time and is lower than full end-to-end bandwidth. The boxes demarcate the quartiles of transfer rate to each facility, collected from 390 transfer tasks where at least 10 GB were transmitted.}
    \label{fig:globus}
     \vspace{-1.5em}
\end{figure}

Figure \ref{fig:globus} shows the effective cross-facility Globus transfer rates achieved in this work, summarized over a sample of 390 transfer tasks from the APS in which at least 10 GB of data were transmitted.  We observed data rates from the APS to ALCF-Theta data transfer nodes were significantly lower than those to the OLCF and NERSC facilities, and needs further investigation.  To adapt to various bandwidths and workload sizes, the Balsam site's transfer module provides an adjustable \emph{transfer batch size} parameter, which controls the maximum number of files per transfer task.  When files are small relative to the available bandwidth, batching files is essential to leveraging the concurrency and pipelining capabilities of GridFTP processes underlying the Globus Transfer service \cite{Yildirim2012}.  On the other hand, larger transfer batch sizes yield diminishing returns on effective speed, increase the duration of transfer tasks, and may decrease opportunities for overlapping computation with smaller concurrent transfer tasks.  

\begin{figure}
    \centering
    \includegraphics[width=0.4\textwidth]{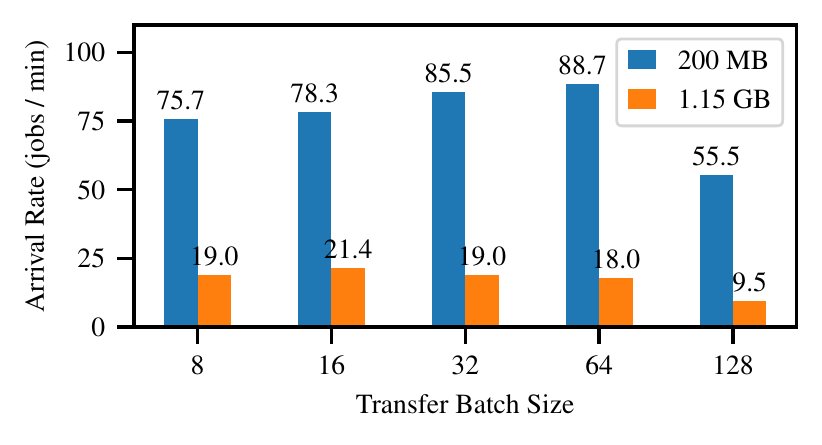}
     \vspace{-1.5em}
    \caption{APS dataset arrival rate for matrix diagonalization benchmark on Theta, averaged over 128 Balsam Jobs.  The Balsam site initiates up to three concurrent Globus transfers with a varying number of files per transfer (transfer batch size).  For batch sizes of 64 and 128, the transfer is accomplished in only two ($128/64$) or one ($128/128$) Globus transfer tasks, respectively.  The diminished rate at batch size 128 shows that at least two concurrent transfer tasks are needed to utilize the available bandwidth.}
    \label{fig:transfer-bs}
     \vspace{-1.5em}
\end{figure}

These tradeoffs are explored for the small and large MD datasets between the APS and a Theta Balsam site in Figure \ref{fig:transfer-bs}.  As expected for the smaller dataset (each file is 200 MB), the aggregate job stage-in rate steadily improves with batch size, but then drops when batch size is set equal to the total workload size of 128 tasks.  We attribute this behavior to the limited default concurrency of 4 GridFTP processes \emph{per transfer task}, which is circumvented by allowing Balsam to manage multiple smaller transfer tasks concurrently.  For the larger MD dataset (each 1.15 GB), the optimal transfer batch size appears closest to 16 files.

\subsection{Autoscaling and fault tolerance}

\begin{figure}
    \centering
    \includegraphics[width=0.45\textwidth]{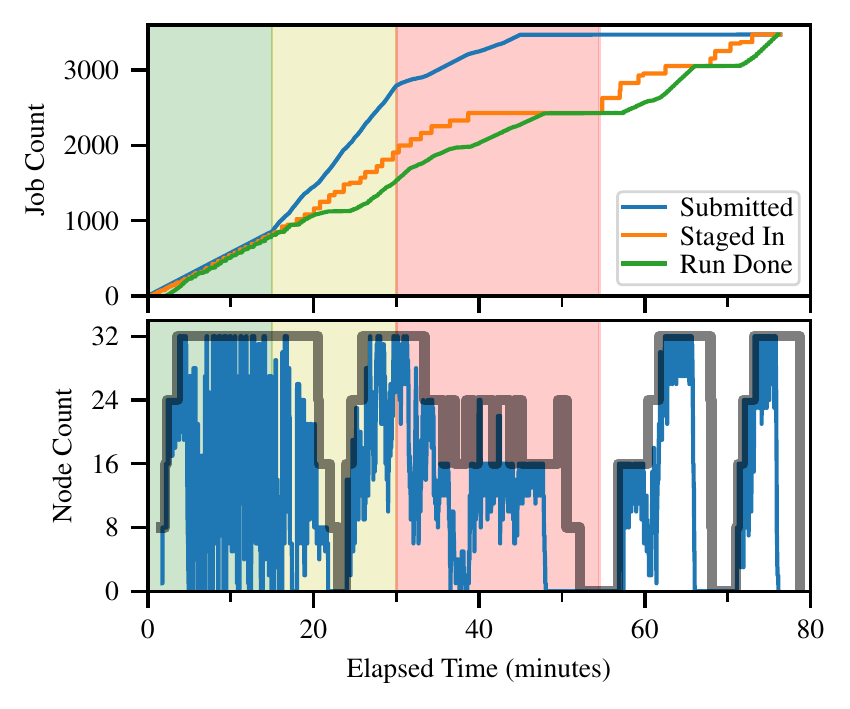}
     \vspace{-1.5em}
    \caption{Elastic scaling during a Balsam stress test using the APS $\leftrightarrow$ Theta matrix diagonalization benchmark with 200 MB dataset.  Three traces in the top panel show the count of jobs submitted to the API, jobs for which the input dataset has been staged, and completed application runs.  The bottom timeline shows the count of elastically-provisioned Theta compute nodes (gray) and count of running analysis tasks (blue).  During the first 15 minutes (green region), jobs are submitted at a steady rate of 1.0 jobs/second.  In the second 15 minutes (yellow region), jobs submitted at a rate of 3.0 jobs/second cause the task backlog to grow, because datasets arrive faster than can be handled with the maximum elastic scaling capacity (capped at 32 nodes for this experiment).  In the third phase (red region), a randomly-chosen Balsam launcher is terminated every two minutes.  In the final stage of the run, Balsam recovers from the faults to fully process the backlog of interrupted jobs. }
    \label{fig:backlog}
     \vspace{-1.5em}
\end{figure}

The Balsam site provides an \emph{elastic queueing} capability, which automates HPC queuing to provision resources in response to time-varying workloads.  When job injection rates exceed the maximum attainable compute nodes in a given duration, Balsam  durably accommodates the growing backlog and guarantees eventual execution in a fashion that is tolerant to faults at the Balsam service, site, launcher pilot job, or individual task level.  Figure \ref{fig:backlog} shows the throughput and node utilization timelines of an 80-minute experiment between Theta and the APS designed to stress test these Balsam capabilities.  

In the first phase (green background), a relatively slow API submission rate of 1.0 small MD job/second ensures that the completed application run count (green curve) closely follows the injected job count (blue curve) with an adequately-controlled latency. The gray trace in the bottom panel shows the number of available compute nodes quickly increasing to 32, obtained via four resource allocations in 8-node increments.  As an aside, we mention that the node count and duration of batch jobs used to provision blocks of HPC resources is adjustable within the Balsam site configuration. In fact, the elastic queuing module may opportunistically size jobs to fill idle resource gaps, using information provided by the local batch scheduler.  In this test, however, we fix the resource block size at 8 nodes and a 20 minutes maximum wall-clock time.  

In the second phase (yellow background), the API submission rate triples from 1.0 to 3.0 small MD jobs/second.  Meanwhile, the available node count briefly drops to 0, as the 20 minute limit expires for each of the initially allocated resource blocks.  Application throughput recovers after the elastic queuing module provisions additional resources, but the backlog between submitted and processed tasks quickly grows larger.  In the third phase (red background), we simulate faults in pilot job execution by \emph{randomly terminating} one of the executing batch jobs every two minutes, thereby causing up to 8 MD tasks to timeout.  The periodically killed jobs, coupled with large scheduler startup delays on Theta, causes the number of available nodes to fluctuate between 16 and 24.  Additionally, the Balsam site experiences a stall in Globus stage-ins, which exhausts the runnable task backlog and causes the launchers to time-out on idling.  Eventually, in the fourth and final region, the adverse conditions are lifted, and Balsam processes the full backlog of tasks across several additional resource allocations.  We emphasize that no tasks are lost in this process, as the Balsam service durably tracks task states in its relational database and monitors the heartbeat from launchers to recover tasks from ungracefully-terminated sessions.  These features enable us to robustly support bursty experimental workloads while providing a fault-tolerant computational infrastructure to the real-time experiments.

\subsection{Simultaneous execution on geographically distributed supercomputers}

  We evaluate the efficacy of Balsam in distributing XPCS (section \ref{sec:apps}) workloads from one or both of the APS and ALS national X-ray facilities on to the Theta, Cori and Summit supercomputing systems in real time. By starting Balsam sites on the login (or gateway or service) nodes of three supercomputers, a user transforms these independent facilities into a federated infrastructure with uniform APIs and a Python SDK to securely orchestrate the life-cycle of geographically-distributed HPC workflows.

\begin{figure}
    \centering
    \includegraphics[width=0.4\textwidth]{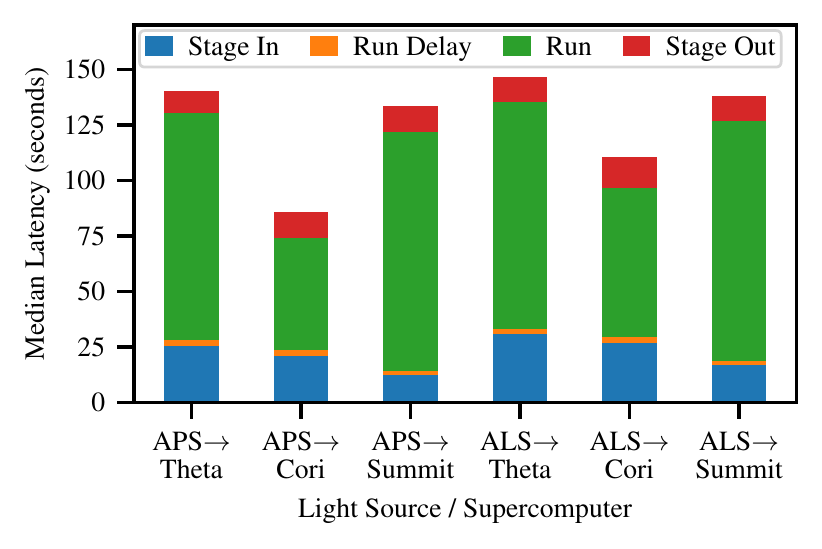}
      \vspace{-1.5em}
    \caption{Contribution of stages in the Balsam Job life-cycle to total XPCS analysis latency.  With at most one 878 MB dataset in flight
    to each compute facility, the bars reflect round-trip time to solution in the absence of pipelining or batching data transfers.  The green \emph{Run} segments represent the median runtime of the XPCS-Eigen \texttt{corr} calculation. The blue and red segments represent overheads in Globus data transfers.  Orange segment represents the delay in Balsam application startup after input dataset has been staged.}
    \label{fig:latency}
      \vspace{-1.5em}
\end{figure}

Figure \ref{fig:latency} shows the median time spent in each stage of a remote Balsam XPCS \texttt{corr} analysis task execution with 878 MB dataset.  The overall time-to-solution, from task submission to returned calculation results, ranges from 86 seconds (APS$\leftrightarrow$Cori) to a maximum of 150 seconds (ALS$\leftrightarrow$Theta). Measurements were performed with an allocation of 32 nodes on each supercomputer and without pipelining input data transfers (i.e.\ a maximum of one GridFTP transfer in flight between a particular light source and compute facility). These conditions minimize latencies due to Globus transfer task queueing and application startup time. From Figure \ref{fig:latency} it is clear that data transfer times dominate the remote execution overheads consisting of stage in, run delay and stage out. However, these overheads are within the constraints needed for near-real-time execution. We emphasize that there is significant room for network performance optimizations and opportunities to interface the Balsam transfer service with performant alternatives to disk-to-disk transfers.  Moreover, the overhead in Balsam application launch itself is consistently in the range of 1 to 2 seconds (1\% to 3\% of XPCS runtime, depending on the platform).

\begin{figure*}
    \centering
    \includegraphics[width=0.75\textwidth]{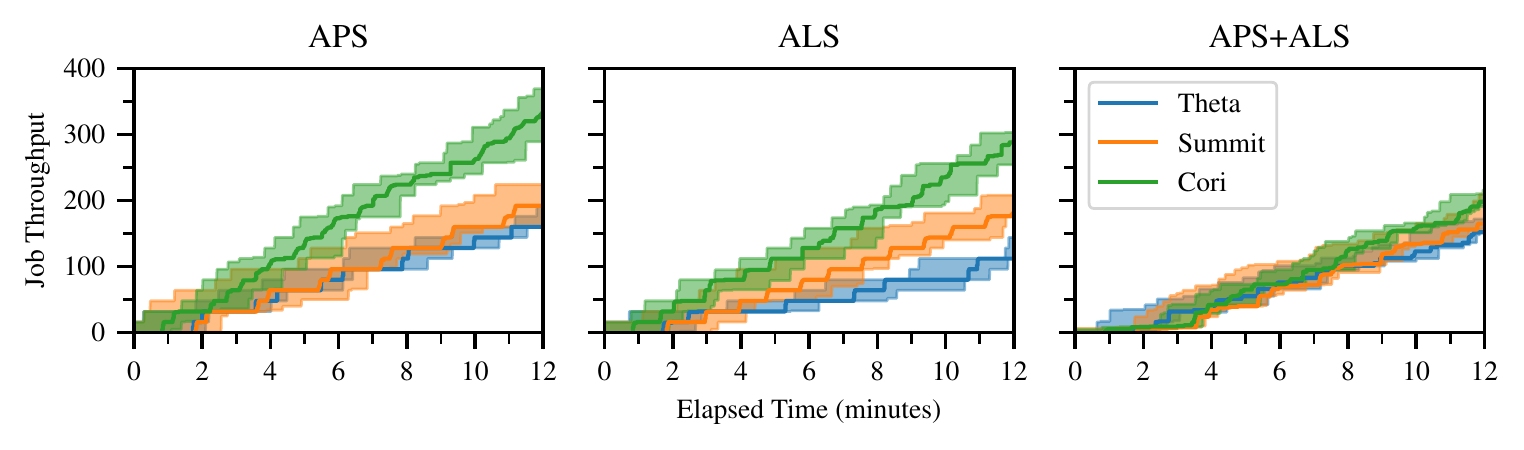}
    \vspace{-1.5em}
    \caption{Simultaneous throughput of XPCS analysis on Theta (blue), Summit (orange), and Cori (green) Balsam sites.  The panels (left to right) correspond to experiments in which the 878 MB XPCS dataset is transferred from either the APS, ALS, or both light sources.  A constant backlog of 32 tasks was maintained with the Balsam API.  The \emph{top} boundary of each shaded region shows the number of staged-in datasets.  The \emph{center} line shows the number of completed application runs.  The \emph{bottom} boundary of the shaded regions shows the number of staged-out result sets.  When the center line approaches the top of the shaded region, the system is network I/O-bound: that is, compute nodes become idle waiting for new datasets to arrive.}
    \label{fig:steady-thru}
    \vspace{-1em}
\end{figure*}

Figure \ref{fig:steady-thru} charts the simultaneous throughput of XPCS analyses executing concurrently on 32 compute nodes of each system:  Theta, Cori, and Summit.  The three panels correspond to separate experiments in which (1) XPCS is invoked from APS datasets alone, (2) XPCS is invoked from ALS datasets alone, or (3) XPCS is invoked with a randomly selected dataset from either APS or ALS. To measure the attainable throughput in each scenario, the clients varied submission rate to maintain a \emph{steady-state backlog} of 32 XPCS tasks: that is, the sum of submitted and staged-in (but not yet running) tasks fluctuates near 32 for each system.    The Balsam site transfer batch size was set to a maximum of 32 files per transfer, and each site initiated up to 5 concurrent transfer tasks.  On the Globus service backend, this implies up to 3 \emph{active} and 12  \emph{queued} transfer tasks at any moment in time.  In practice, the stochastic arrival rates and interwoven stage-in and result stage-out transfers reduce the file count per transfer task.  

One immediately observes a consistent ordering in throughput increasing from Theta $\rightarrow$ Summit $\rightarrow$ Cori, regardless of the client light source.  The markedly improved throughput on Cori relative to other systems is mainly due to reduced application runtime, which is clearly evident from Figure \ref{fig:latency}.  While the throughput of Theta and Summit are roughly on-par, Summit narrowly though consistently outperforms Theta due to higher effective stage-in throughput bandwidth.  This is evident in the \emph{top boundary} of the orange and blue regions of Figure \ref{fig:steady-thru}, which show the arrival throughput of staged-in datasets.  Indeed, the average XPCS arrival rates calculated from the APS experiment (left panel) are 16.0 datasets/minute for Theta, 19.6 datasets/minute for Summit, and 29.6 datasets/minute for Cori.  We also notice that Cori's best data arrival rate is inconsistent with the slower median stage in time of Figure \ref{fig:latency}. Because those durations were measured without batching file transfers, this underscores the significance of GridFTP pipelining and concurrency on  the Cori data transfer nodes.

Continuing our analysis of Figure \ref{fig:steady-thru}, the \emph{center line} in each colored region shows throughput of completed application runs, which naturally lags behind the staged-in datasets by the XPCS application's running duration. Similarly, the \emph{bottom boundary} of each colored region shows the number of completed result transfers back to the client; the gap between center line and bottom of the region represents ``Stage Out'' portion of each pipeline.  The XPCS results comprise only the HDF file, which is modified in-place during computation and (at least) an order-of-magnitude smaller than  \texttt{IMM} frame data.  Therefore, result transfers tend to track application completion more closely, which is especially noticeable on Summit.

\begin{figure*}
    \centering
    \includegraphics[width=0.75\textwidth]{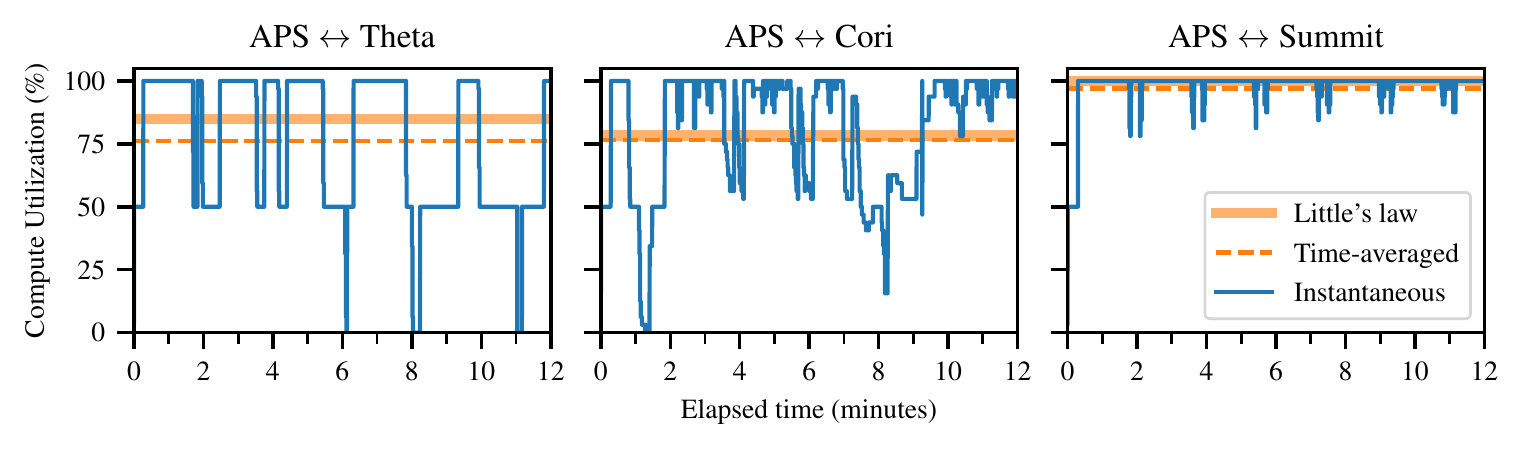}
    \vspace{-2em}
    \caption{Compute node utilization on each Balsam site corresponding to the APS experiment in Figure \ref{fig:steady-thru}. The solid line shows the instantaneous node utilization, measured via Balsam events as a percentage of the full 32-node reservation on each system.  The dashed line indicates the time-averaged utilization, which should coincide closely with the \emph{expected} utilization derived from the average data arrival rate and application runtime (Little's law). The data arrival rate on Summit is sufficiently high that compute nodes are busy near 100\% of the time.  By contrast, Theta and Cori reach about 75\% utilization due to a bottleneck in effective data transfer rate.}
    \label{fig:steady-util}
    \vspace{-1em}
\end{figure*}

When the center line in a region of Figure \ref{fig:steady-thru} approaches or intersects the top region boundary, the number of \emph{completed} application runs approaches or matches the number of \emph{staged-in} XPCS datasets.  In this scenario, the runnable workload on a supercomputer becomes too small to fill the available 32 nodes: in effect, the system is network I/O-bound.  The compute utilization percentages in Figure \ref{fig:steady-util} make this abundantly clear for the APS experiment.  On Summit, the average arrival rate of 19.6 datasets/minute and 108 second application runtime are sufficiently high to keep all resources busy.  Indeed, for Summit both  instantaneous utilization measured from Balsam event logs (blue curve) and its time-average (dashed line) remain very  close to 100\% ($32/32$ nodes running XPCS analysis).  

We can apply Little's law \cite{Little2008} in this context, an intuitive result from queuing theory which states that the long-run average number $L$ of running XPCS tasks should be equal to the product of average effective dataset arrival rate $\lambda$ and application runtime $W$.  We thus compute the empirical quantity $L = \lambda W$ and plot it in Figure \ref{fig:steady-util} as the semi-transparent thick yellow line.  The close agreement between time-averaged node utilization and the expected value $\lambda W$ helps us to estimate the \emph{potential headroom for improvement of throughput} by increasing data transfer rates ($\lambda$).   For example, even though APS$\leftrightarrow$ Cori achieves better overall throughput than APS $\leftrightarrow$ Summit (Figure \ref{fig:steady-thru}), we see that Cori reaches only about 75 \% of its sustained XPCS processing capacity at 32 nodes, whereas Summit is near 100 \% utilization and therefore compute-bound (of course barring optimizations in application I/O or other acceleration techniques at the task level).  Furthermore, Figure \ref{fig:latency} shows that XPCS application runtime on Theta is comparable to Summit, but Theta has a significantly slower arrival rate. This causes lower average utilization of 76\% on Theta.

Overall, by using Balsam to scale XPCS workloads from the APS across three distributed HPC systems simultaneously, we achieved 4.37-fold increased throughput over the span of a 19-minute run compared to routing workloads to Theta alone (1049 completed tasks in aggregate, versus 240 tasks completed on Theta alone). Similarly, we achieve a 3.28X improvement in throughput over Summit and 2.2X improvement over Cori.  
 In other words, depending on the choice of reference system, we obtain a 73\% to 146\% ``efficiency'' of throughput scaling when moving from 32 nodes on one system to 96 nodes across the three. Of course, these figures should be loosely interpreted due to the significant variation in XPCS application performance on each platform.
This demonstrates the importance of utilizing geographically distributed supercomputing sites to meet the needs of real-time analysis of experiments for improved time to solution. We note that the Globus Transfer default limit of 3 concurrent transfer tasks per user places a significant constraint on peak throughput achievable by a single user.  This issue is particularly evident in the right-most panel of Figure \ref{fig:steady-thru}, where Balsam is managing bi-directional transfers along six routes.  As discussed earlier, the clients in this experiment throttled submission to maintain a \emph{steady backlog} of 32 jobs per site.  Splitting this backlog in half between two remote data sources (APS and ALS) decreases opportunities for batching transfers and reduces overall benefits of pipelining/concurrency in file transfers.   Of course, this is to some extent an artifact of the experimental design: real distributed clients would likely use different Globus user accounts and would \emph{not} throttle task injection to maintain a steady backlog.  Nevertheless, increasing concurrency limits on data transfers could provide significant improvements in global throughput where many Balsam client--execution site endpoint pairs are concerned.  

\begin{figure}
    \centering
    \includegraphics[width=0.35\textwidth]{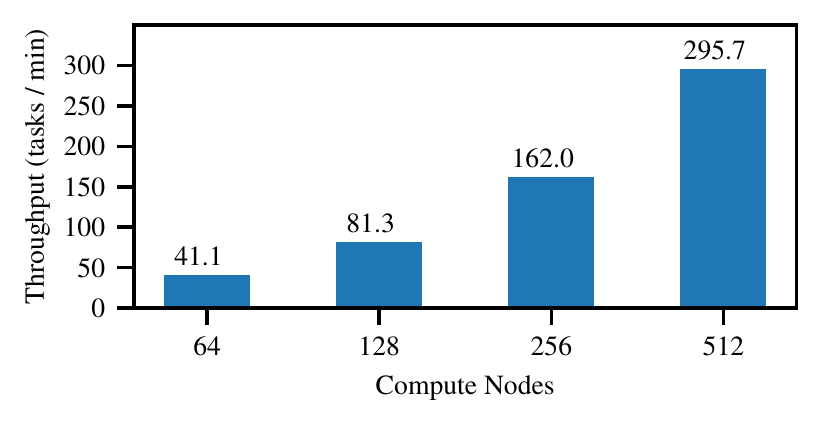}
     \vspace{-2em}
    \caption{ Weak scaling of XPCS benchmark throughput with increasing Balsam launcher job size on Theta.  We consider how throughput of the same XPCS application increases with compute node count when the network data transfer bottleneck is removed. An average of two jobs per Theta node were completed at each node count, with 90\% efficiency at 512 nodes.}
    \vspace{-2em}
   \label{fig:mpi-scaling}
\end{figure}

Although we have established the limiting characteristics of XPCS data transfers on Balsam scalability, one may consider more generally the scalability of Balsam on large-scale HPC systems.  Figure \ref{fig:mpi-scaling} shows the performance profile of the XPCS workload on Theta when wide-area-network data movement is taken out of the equation and the input datasets are read directly from local HPC storage.  As more resources are made available to Balsam launchers, the rate of XPCS processing weak-scales with 90\% efficiency from 64 to 512 Theta nodes.  This efficiency is achieved with the \texttt{mpi} pilot job mode where one \texttt{aprun} process is spawned for each XPCS job.  We anticipate improved scaling efficiency with the \texttt{serial} job mode.  This experiment was performed with a fixed number of XPCS tasks per compute node; however, given the steady task throughput, we expect to obtain similar strong scaling efficiency in reducing time-to-solution for a fixed large batch of analysis tasks.  Finally, launchers obtain runnable jobs from the Balsam Session API, which fetches a variable quantity of \texttt{Jobs} to occupy available resources. Because runnable Jobs are appropriately indexed in the underlying PostgreSQL database, the response time of this endpoint is largely consistent with respect to increasing number of submitted Jobs.  We therefore expect consistent performance of Balsam as the number of backlog tasks grows.

\subsection{Adaptive workload distribution}

The previous section focused on throughputs and latencies in the Balsam distributed system for analysing XPCS data on Theta, Summit, and Cori execution sites.  We now relax the artificial \emph{steady backlog} constraint and instead allow the APS client to inject workloads in a more realistic fashion: jobs are submitted at a \emph{constant average rate}  of 2.0 jobs/second, spread across batch-submission blocks of \emph{16 XPCS jobs every 8.0 seconds}.  This workload represents a plausible experimental workflow, in which 14 GB batches of data (16 datasets $\times$ 878 MB/dataset) are periodically collected and dispatched for XPCS-Eigen analysis remotely.  It also introduces a scheduling problem, wherein the client may naively distribute workloads evenly among available execution sites (round-robin) or leverage the Balsam Job events API to \emph{adaptively} route tasks to the system with shortest backlog, lowest estimated time-to-solution, etc\ldots  Here, we explore a simplified variant of this problem, where XPCS tasks are uniformly-sized and each Balsam site has a constant resource availability of 32 nodes.  

\begin{figure}
    \centering
    \includegraphics[width=0.37\textwidth]{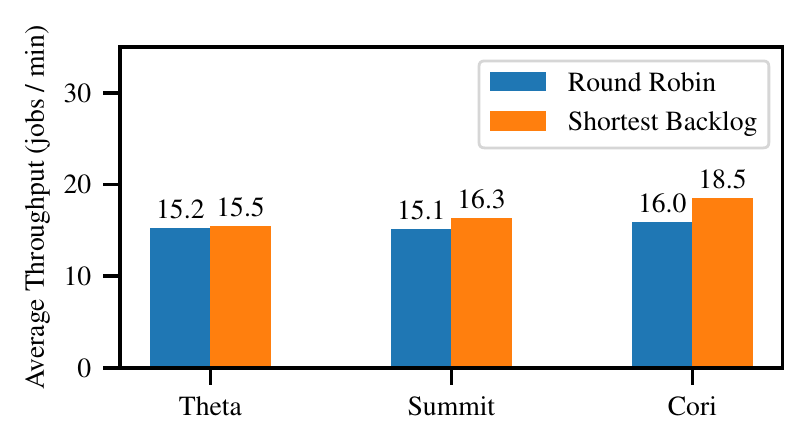}
    \vspace{-1.5em}
    \caption{Throughput of client-driven task distribution strategies for the XPCS benchmark with APS data source. Using round-robin distribution, XPCS jobs are evenly distributed alternating among the three Balsam sites. Using the shortest-backlog strategy, the client uses the Balsam API to adaptively send jobs to the site with the smallest pending workload.  In both scenarios, the client submits a batch of 16 jobs every 8 seconds.}
    \label{fig:rr-vs-sb-thru}
    \vspace{-1em}
\end{figure}

Figure \ref{fig:rr-vs-sb-thru} compares the average throughput of XPCS dataset analysis from the APS when distributing work according to one of two strategies: \emph{round-robin} or \emph{shortest-backlog}.  In the latter strategy, the client uses the Balsam API to poll the number of jobs pending stage-in or execution at each site; the client then submits each XPCS batch to the Balsam site with the shortest backlog in a first-order load-levelling attempt.  We observed a slight 16\% improvement in throughput on Cori when using the shortest-backlog strategy relative to  round-robin, with marginal differences for Theta and Summit.  The submission rate of 2.0 jobs/second or 120 jobs/minute significantly exceeds the aggregate sustained throughput of approximately 58 jobs/minute seen in the left panel of Figure \ref{fig:steady-thru}.  We therefore expect a more significant difference between the distribution strategies when job submission rates are balanced with the attainable system throughput.  

\begin{figure}
    \centering
    \includegraphics[width=0.40\textwidth]{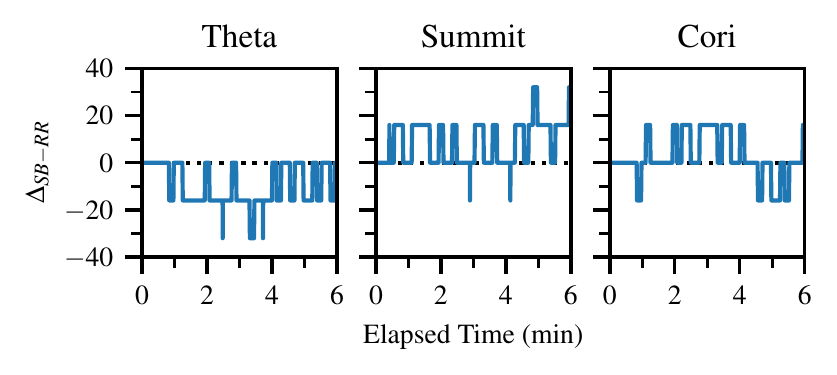}
    \vspace{-1.5em}
    \caption{Difference in per-site job distribution between the considered strategies.  Each curve shows the instantaneous difference in submitted task count between the shortest-backlog and round-robin strategies ($\Delta_{SB-RR}$) over the span of six minutes of continuous submission. A negative value indicates that \emph{fewer} jobs were submitted at the same point in time during the round-robin run, relative to the shortest-backlog run.  }
    \label{fig:rr-vs-sb}
    \vspace{-1.5em}
\end{figure}

\begin{figure}
    \centering
    \includegraphics[width=0.40\textwidth]{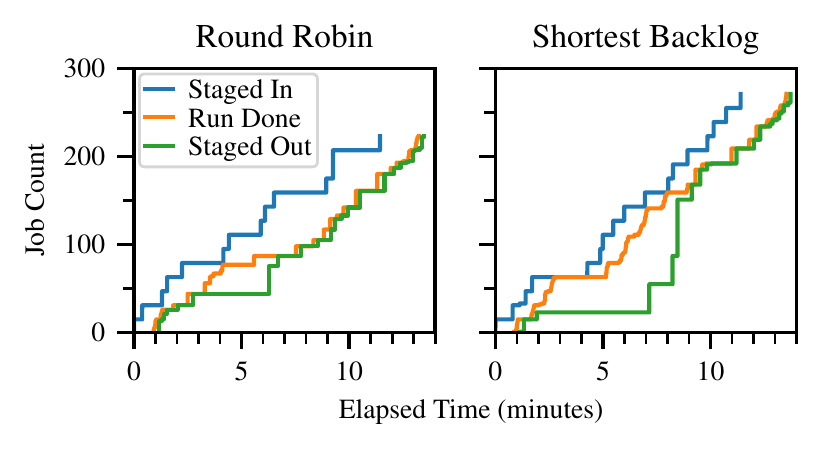}
    \vspace{-1.5em}
    \caption{Throughput of data staging and application runs on Cori, compared between round-robin and shortest-backlog strategies.  More frequent task submissions to Cori in the adaptive shortest-backlog strategy contribute to 16\% higher throughput relative to round-robin strategy. }
    \label{fig:rr-vs-sb-cori}
    \vspace{-1.5em}
\end{figure}

Nevertheless, Figure \ref{fig:rr-vs-sb} shows that with the shortest-backlog strategy, the APS preferentially submits more workloads to Summit and Cori, and fewer workloads to Theta, which tends to accumulate backlog more quickly owing to slower data transfer rates.  This results in overall increased data arrival rates and higher throughput on Cori (Figure \ref{fig:rr-vs-sb-cori}), despite overloaded pipelines on all systems.

\section{Conclusion}

In this work, we describe the Balsam workflow framework to enable wide-area, multi-tenant, distributed execution of analysis jobs on DOE supercomputers. A key development is the introduction of a central Balsam service that manages job distribution, fronted by a programmatic interface. Clients of this service use the API to inject and monitor jobs from anywhere. 
A second key development is the Balsam site, which consumes jobs from the Balsam service and interacts directly with the scheduling infrastructure of a cluster. Together, these two form a distributed, multi-resource execution landscape. 

The power of this approach was demonstrated using the Balsam API to inject analysis jobs from two light source facilities to the Balsam service, targeting simultaneous execution on three DOE supercomputers. By strategically combining data transfer bundling and controlled job execution, Balsam achieves higher throughput than direct-scheduled local jobs and effectively balances job distribution across sites to simultaneously utilize multiple remote supercomputers. Utilizing three geographically distributed supercomputers concurrently, we achieve a 4.37-fold increased throughput compared to routing light source workloads to Theta. 
Other key results include that Balsam is a user-domain solution that can be deployed without administrative support, while honoring the security constraints of facilities. 
\bibliographystyle{ACM-Reference-Format}
\bibliography{main}


\end{document}